\documentclass{IEEEcsmag}

\usepackage[colorlinks,urlcolor=blue,linkcolor=blue,citecolor=blue]{hyperref}
\expandafter\def\expandafter\UrlBreaks\expandafter{\UrlBreaks\do\/\do\*\do\-\do\~\do\'\do\"\do\-}

\jvol{XX}
\jnum{XX}
\paper{XX}
\jmonth{Month}
\jname{IEEE Software}
\jtitle{IEEE Software}
\pubyear{2023}

\usepackage[
	autocite=superscript,
	sorting=none,
	backend=biber,
	bibencoding=utf8,
	isbn=false,
	url=false,
	doi=true,
]{biblatex}
\defbibenvironment{bibliography}{\enumerate}{\endenumerate}{\item}

\usepackage{wrapfig}

\usepackage{charter}
\usepackage[charter]{mathdesign}

\usepackage[framemethod=TikZ]{mdframed}

\usepackage{booktabs}
\usepackage{multirow}

\usepackage{xcolor}

\usepackage{tikz}
\usetikzlibrary{automata, positioning, arrows, decorations.pathreplacing, calc, chains, shapes.misc, backgrounds, fit, patterns, fadings, shadows.blur}
\usepackage{pgfplots}
\pgfplotsset{compat=newest}
\usepgfplotslibrary{fillbetween, dateplot}

\usepackage{ifthen}

\pgfdeclarelayer{background}
\pgfsetlayers{background, main}

\definecolor{ACMYellow}{RGB}{255, 214, 0}
\definecolor{ACMOrange}{RGB}{252, 146, 0}
\definecolor{ACMRed}{RGB}{253, 27, 20}
\definecolor{ACMLightBlue}{RGB}{131, 206, 226}
\definecolor{ACMGreen}{RGB}{166, 188, 9}
\definecolor{ACMPurple}{RGB}{101, 1, 107}
\definecolor{ACMDarkBlue}{RGB}{9, 53, 122}

\newcommand*\circled[1]{\tikz[baseline=(char.base)]{
            \node[shape=circle,draw,inner sep=2pt, thick] (char) {#1};}}

\usepackage{xparse}
 
\let\oldsection\section
\RenewDocumentCommand{\section}{s o m}{%
  \IfBooleanTF{#1}
    {\oldsection*{\uppercase{#3}}}
    {\IfValueTF{#2}
       {\oldsection[\uppercase{#2}]{\uppercase{#3}}}
       {\oldsection{\uppercase{#3}}}
    }%
}

\begin{document}


\sptitle{Article Type: Original research}

\title{Taxing Collaborative Software Engineering}

\author{Michael Dorner}
\affil{Blekinge Institute of Technology, Sweden}

\author{Maximilian Capraro}
\affil{Kolabri, Germany}

\author{Oliver Treidler}
\affil{Kolabri, Germany}

\author{Tom-Eric Kunz}
\affil{Kolabri, Germany}

\author{Darja Smite}
\affil{Blekinge Institute of Technology, Sweden}

\author{Ehsan Zabardast}
\affil{Blekinge Institute of Technology, Sweden}

\author{Daniel Mendez}
\affil{Blekinge Institute of Technology, Sweden and fortiss, Germany}

\author{Krzysztof Wnuk}
\affil{Blekinge Institute of Technology, Sweden}

\begin{abstract}
The engineering of complex software systems is often the result of a highly collaborative effort. However, collaboration within a multinational enterprise has an overlooked legal implication when developers collaborate across national borders: It is taxable. 
In this article, we discuss the unsolved problem of taxing collaborative software engineering across borders. We (1) introduce the reader to the basic principle of international taxation, (2) identify three main challenges for taxing collaborative software engineering making it a software engineering problem, and (3) estimate the industrial significance of cross-border collaboration in modern software engineering by measuring cross-border code reviews at a multinational software company. 
\end{abstract}

\maketitle

\section{Introduction}

\begin{quote}
``He's spending a year dead for tax reasons.''\begin{flushright}\small{-- Douglas Adams, \textit{The Hitchhiker's Guide to the Galaxy}}\end{flushright}
\end{quote}

\chapteri{M}odern software systems are too large, too complex, and evolving too fast for single developers to oversee. Therefore, software engineering has become highly collaborative. Further, software development is often a joint effort of individuals and teams collaborating across borders, especially in multinational companies with their subsidiaries spread around the globe.\autocites{Herbsleb2001} However, \textbf{collaboration has a legal implication if individuals collaborate across borders: the profits from those cross-border collaborations become taxable}. 

In this article, we describe the complexity of applying the established international taxation standards required and enforced by national tax authorities in the context of modern software engineering with its distributed and fine-grained collaboration crossing borders. We start with a gentle introduction to international standards in multinational taxation and its basic \emph{arm's length principle} for software engineers. We then discuss the challenges of taxing collaborative software engineering and illustrate the industrial significance of cross-border collaboration in an industrial case, namely code review. 

Taxation in software industry has been debated for many decades.\autocite{OECD2015} The problem with taxing the final result of software engineering, the software product or service, for example, has shown to be challenging to tackle and is still subject to ongoing and broad and broad discussion.\autocite{Olbert2017} Here, we extend the debate to software engineering, the way the software products and services are being developed, which has not yet been covered. Our goal is to raise a debate and draw attention to this problem among a software engineering audience. For in-depth information on basic transfer pricing concepts including standard methods and tax compliance requirements, we recommend the interested reader further readings.\autocite{Treidler2020}

\section{A gentle introduction to taxing multinational enterprise for software engineers}

Consider \textit{devnullsoft Group}, a multinational enterprise that develops and sells software-intensive product, has two legal entities: \textit{devnullsoft GmbH} in Germany and its subsidiary \textit{devnullsoft AB} in Sweden. The German development team employed by \textit{devnullsoft GmbH} develops the software-intensive product jointly with the Swedish development team employed by the Swedish subsidiary \textit{devnullsoft AB}. The German \textit{devnullsoft GmbH} sells this resulting product to customers. 

Without any further consideration, solely the German \textit{devnullsoft GmbH} generates profits, which are then fully taxed in Germany according to German law. The Swedish tax authorities are left out in the cold because \textit{devnullsoft AB} has no share of the profit that could be taxed in Sweden, although \textit{devnullsoft AB} contributed significantly to the product through code contributions, code reviews, bug reports, tests, architectural decisions, or other contributions that made the success of the software possible

To avoid this scenario and to provide a common ground for international taxation, reducing uncertainty for multinational enterprises, and preventing tax avoidance through profit shifting, nearly all countries in the world agreed on and implemented the so-called \emph{arm's length principle} as defined in the \emph{OECD Transfer Pricing Guidelines for Multinational Enterprises and Tax Administrations}.\autocite{OECD2022}

The \emph{arm's length principle} is the guiding principle and the de-facto standard for the taxation of multinational enterprises that requires associated enterprises to operate as if not associated and regular participants in the market from a taxation perspective. This principle ensures that transfer prices between associated companies of multinational enterprises are established on a market value basis and not misused for profit shifts from high to low tax regions. 

To comply with the arm's length principle, \textit{devnullsoft GmbH} in Germany and \textit{devnullsoft AB} in Sweden need to operate from a taxation perspective as if they were not associated. Since a regular participant in the market would not provide code contributions, code reviews, tests, or architectural designs free or other contributions of charge to a closed-source software project, \textit{devnullsoft GmbH} in Germany needs to pay for the received contributions, the so-called \emph{transfer price}. 

Transfer prices are the prices at which an enterprise transfers physical goods and intangibles or provides services to associated enterprises. Since software is intangible itself, the transfer of intangibles like source code, code reviews, bug reports, etc., is our focal point. This transfer price guarantees that \textit{devnullsoft AB} gets its share of the profit, which then can be taxed by the Swedish tax authorities.

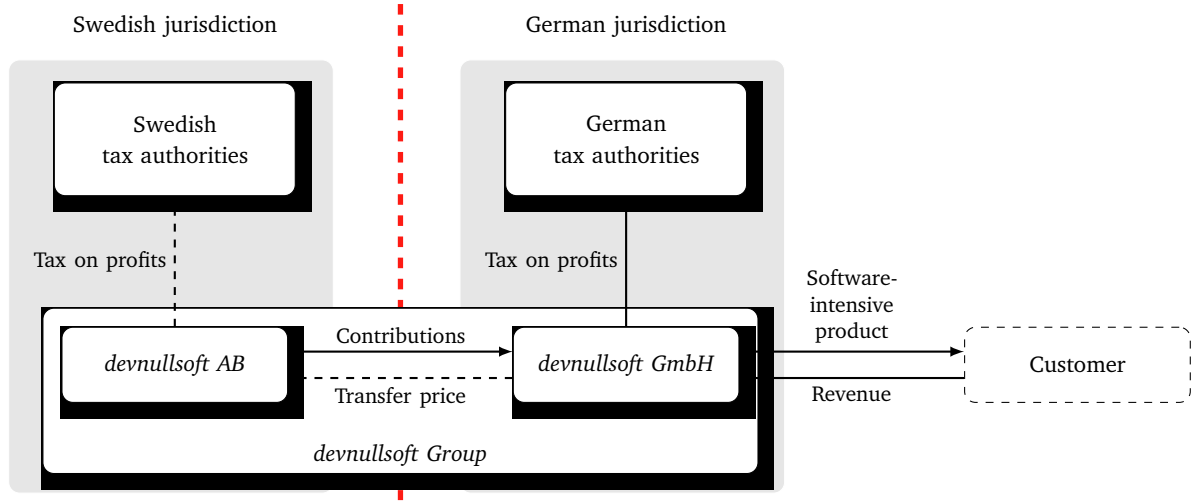
\begin{figure*}\centering
\begin{tikzpicture}[x=6cm, y=3cm,
taxauthority/.style={rounded corners=1ex, draw=black, fill=ACMRed!0, minimum width=3cm, minimum height=1.5cm, fill=white, text width=3cm, align=center, blur shadow={shadow opacity=25, shadow xshift=1mm, shadow yshift=-1mm}},
company/.style={draw=black, minimum width=3cm, minimum height=1cm, fill=white, rounded corners=1ex, blur shadow={shadow opacity=25, shadow xshift=1mm, shadow yshift=-1mm}},
flow/.style={-latex, thick},
]

\node[taxauthority] (swedishtax) at (0,1) {Swedish\\tax authorities};
\node[taxauthority] (germantax) at (1,1) {German\\tax authorities};
\node[company] (swedishcompany) at (0,0) {\textit{devnullsoft AB}};
\node[company] (germancompany) at (1,0) {\textit{devnullsoft GmbH}};
\node[draw=black, minimum width=3cm, minimum height=1cm, fill=white, dashed, rounded corners=1ex] (customer) at (2,0) {Customer};


\path[flow] (swedishcompany) edge[bend right=0, dashed] node[font=\small, left] {Tax on profits} (swedishtax);
\path[flow] (germancompany) edge[bend right=0] node[font=\small, left] {Tax on profits} (germantax);

\path[flow] ([yshift=5pt]swedishcompany.east) edge[bend left=0] node[font=\small, above, align=center, text width=4.75cm] {Contributions} ([yshift=5pt]germancompany.west);
\path[flow] ([yshift=-5pt]germancompany.west) edge[dashed, bend left=0] node[font=\small, below] {Transfer price} ([yshift=-5pt]swedishcompany.east);

\path[flow, bend right=0] ([yshift=5pt]germancompany.east) edge node[font=\small, above, text width=1.75cm, align=center] {Software-intensive product} ([yshift=5pt]customer.west);
\path[flow, bend right=0] ([yshift=-5pt]customer.west) edge[bend right=0] node[font=\small, below] {Revenue} ([yshift=-5pt]germancompany.east);

\begin{scope}[on background layer]	

\draw[line width=2pt, dash pattern=on 4pt off 4pt, ACMRed] (0.5, 1.6) -- (0.5, -0.6);
	
\draw[black!0, fill=black!10, rounded corners=1ex] ($ (swedishtax.north west) + (-0.1,0.1)$) rectangle ($ (swedishcompany.south east) + (0.1,-0.4)$);
\node[above=0.5cm of swedishtax] {Swedish jurisdiction};
	
\draw[black!0, fill=black!10, rounded corners=1ex] ($ (germantax.north west) + (-0.1,0.1)$) rectangle ($ (germancompany.south east) + (0.1,-0.4)$);
\node[above=0.5cm of germantax] {German jurisdiction};

\coordinate (group) at (0.5,-0.4);
\node[draw, fit=(swedishcompany)(germancompany)(group), inner sep=7pt, black, fill=black!0, rounded corners=1ex, blur shadow={shadow opacity=25, shadow xshift=1mm, shadow yshift=-1mm}] {};
\node at (group.south) {\textit{devnullsoft Group}};

\end{scope}
\end{tikzpicture}
\caption{A schematic overview of the necessity and mechanics of transfer pricing in a multinational enterprise (\textit{devnullsoft Group}) with two associated software companies (\textit{devnullsoft AB} in Sweden and \textit{devnullsoft GmbH} in Germany): Without considering a market-based compensation, the so-called \emph{transfer price}, \textit{devnullsoft AB} has no share on the profits which could be taxed by the Swedish tax authorities; all profits are with \textit{devnullsoft GmbH} and, therefore, all taxes stay within Germany.}
\label{fig:overview}
\end{figure*}

In Figure~\ref{fig:overview}, we provide a schematic overview of transfer pricing between the two associated software companies from our example. Although \textit{devnullsoft AB} contributed significantly to the software-intensive product, without a transfer price, \textit{devnullsoft AB} has no share of profits; all profits are fully taxable in Germany only. However, if \textit{devnullsoft GmbH} in Germany pays a transfer price reflecting the value for the services and intangible properties received from its Swedish associated enterprise, \textit{devnullsoft AB} realizes profits which then are taxable in Sweden. 

In our case, the \textit{devnullsoft Group} does not artificially shift profits to a tax haven. Yet, one can easily imagine that neglecting to charge arm’s length prices can be intentionally misused for profit-shifting. Therefore, the OECD guidelines permit tax authorities like the Swedish tax authority to adjust the transfer price where the prices charged are outside an arm's length range. Such an adjustment will carry interest and might be coupled with penalties. In the wake of the OECD's \emph{Base Erosion and Profit Shifting (BEPS) Project}\footnote{\url{https://www.oecd.org/tax/beps/}}, the regulatory framework has become considerably stricter at an international and national level. As a result, tax authorities can demand more comprehensive information to detect misalignments and enforce tax adjustments. From the companies' perspective, its software development may be---intentionally or unintentionally---non-compliant and face the risk of being legally prosecuted. 

\section{Challenges}\label{sec:challenges}

So, what are transfer prices for collaborative software engineering that comply with this arm's length principle? Determining a market price for intangibles is inherently difficult and is reflected in a broad price range. Collaborative software engineering, however, scales the problem of a transfer-price determination to a new level of complexity because the reality of modern software engineering is significantly more complex than our introductory example above may suggest. Since transfer price regulations apply to a much broader definition of intangibles compared to accounting standards, the latter can not be used as a reliable measure of value for transfer pricing purposes.\autocite{OECD2022}

\begin{figure*}\centering
\begin{tikzpicture}

\begin{scope}[xshift=-2cm, start chain=going below, node distance=9mm]
    \foreach \d in {1,2,3,4}
      \node[on chain, draw, circle, fill=white] (d\d) {$d_\d$};
    \end{scope}

    \begin{scope}[xshift=10cm, start chain=going below, node distance=5mm]
    \foreach \d in {5, 6, ..., 9}
      \node[on chain, draw, circle, fill=white] (d\d) {$d_\d$};

\end{scope}

\begin{scope}[yshift=-5.5cm, on background layer]

\node[draw=black!50, thick, rounded corners=1em, fit={(-3, -0.5) (3.25, 7.5)}, rectangle, inner sep=0mm, label={[yshift=-7mm]above:devnullsoft AB}, fill=black!5] (ab) {};
\node[draw=black!50, thick, rounded corners=1em, fit={(3.75, -0.5) (11, 7.5)}, rectangle, inner sep=0mm, label={[yshift=-7mm]above:devnullsoft GmbH}, fill=black!5] (gmbh) {};


\draw[line width=2pt, dash pattern=on 4pt off 4pt, ACMRed] (3.5, 8) -- (3.5, -1);

\draw[fill=white] (0,0) grid (3,6) rectangle (0,0);
\draw[fill=white] (4,0) grid (8,6) rectangle (4,0);

\foreach \x in {0, 1, ..., 7} 
	\foreach \y in {0, 1, ..., 5} {
		\node[circle, shift={(0.5,0.5)}, draw=none, inner sep=2mm, outer sep=0pt] at (\x, \y) (n\x\y) {};
	}

\node[rectangle, minimum size=1cm, pattern=north west lines] (component) at (n22) {};
\node[rectangle, minimum size=1cm, pattern=north west lines] (reusedcomponent) at (n40) {};

\draw[decorate,decoration={brace,amplitude=10pt,mirror,raise=4pt},yshift=0pt]
(0,-0.5) -- (8, -0.5) node [draw=none, midway, yshift=-0.8cm] {Software system};

\foreach \d/\x/\y/\b/\c in {3/2/5/left/ACMDarkBlue, 2/1/5/left/ACMDarkBlue, 1/1/0/left/ACMDarkBlue, 3/2/3/left/ACMDarkBlue, 1/0/5/left/ACMDarkBlue, 1/2/0/left/ACMDarkBlue, 3/0/0/left/ACMDarkBlue, 1/1/3/left/ACMDarkBlue, 3/0/2/left/ACMDarkBlue, 4/0/4/left/ACMDarkBlue, 1/0/4/left/ACMDarkBlue, 3/2/1/left/ACMDarkBlue, 2/7/4/left/ACMOrange, 2/7/0/left/ACMOrange, 5/6/1/right/ACMDarkBlue, 8/4/3/right/ACMDarkBlue, 9/4/2/right/ACMDarkBlue, 7/5/3/right/ACMDarkBlue, 7/7/4/right/ACMDarkBlue, 9/5/5/right/ACMDarkBlue, 9/5/3/right/ACMDarkBlue, 8/7/5/right/ACMDarkBlue, 6/6/0/right/ACMDarkBlue, 8/5/1/right/ACMDarkBlue, 5/4/4/right/ACMDarkBlue, 8/5/1/right/ACMDarkBlue, 7/7/0/right/ACMDarkBlue, 5/5/2/right/ACMDarkBlue, 5/7/3/right/ACMDarkBlue, 7/6/2/right/ACMDarkBlue, 6/0/5/right/ACMOrange, 6/1/3/right/ACMOrange, 5/1/4/right/ACMOrange, 8/2/3/right/ACMOrange, 8/4/0/right/ACMDarkBlue, 3/4/0/left/ACMOrange} \draw[-latex, thick, color=\c] (d\d) to[bend \b] (n\x\y);

\draw[-latex, color=ACMOrange, thick] (d4) to[bend left] (n40);
\draw[-latex, color=ACMDarkBlue, thick] (d9) to[bend right] (n40);

\draw[-latex, color=ACMOrange, thick] (d1) to[bend left] node[midway, circle, draw, ACMRed, minimum size=0.5cm] (transaction) {} (n65);
\node[circle, draw, thick, ACMRed, minimum size=1.5cm] (account) at (n40) {};
\node[circle, draw, thick, ACMRed, minimum size=0.5cm] (tracking) at (3.5, 7.5) {};

\end{scope}

\begin{scope}[xshift=-2cm, yshift=-9cm, y=0.5cm, label distance=5mm]
\node[draw, circle, anchor=east, label=right:Developer, minimum size=1em] at (0,4) {};
\node[draw, rectangle, anchor=east, label=right:Software component, minimum size=1em] at (0,3) {};
\node[draw, rectangle, anchor=east, pattern=north west lines, label=right:Reused software component, minimum size=1em] at (0,2) {};
\draw[-latex, thick, ACMOrange] (-1, 1) -- (0, 1) node[black, label=right:Taxable transaction, anchor=east] {};
\draw[-latex, thick, ACMDarkBlue] (-1, 0) -- (0, 0) node[label=right:Non-taxable transaction, anchor=east] {};
\end{scope}

\draw[ACMOrange, thick, -latex] (reusedcomponent) to[bend right] (component);

\node[ACMRed, every two node part/.style={text width=4.5cm}, align=left, rectangle split, rectangle split horizontal, rectangle split parts=2] (transactiondesc) at (1, 3) {\circled{1}\nodepart{two}What is a taxable transaction\\in software engineering?};
\draw[ACMRed, thick] (transaction) -- (transactiondesc);

\node[ACMRed, every two node part/.style={text width=6cm}, align=left, rectangle split, rectangle split horizontal, rectangle split parts=2] (trackingdesc) at (8, 3) {\circled{2}\nodepart{two}How to track taxable transactions?};
\draw[ACMRed, thick] (tracking) -- (trackingdesc);

\node[ACMRed, every two node part/.style={text width=7cm}, align=left, rectangle split, rectangle split horizontal, rectangle split parts=2] (accountdesc) at (8, -8) {\circled{3}\nodepart{two} How to value taxable transactions?};
\draw[ACMRed, thick] (account) -- (accountdesc);

\end{tikzpicture}
\caption{A schematic overview of collaborative software engineering and three challenges for transfer pricing specific to software engineering.}
\label{fig:challenges}
\end{figure*}
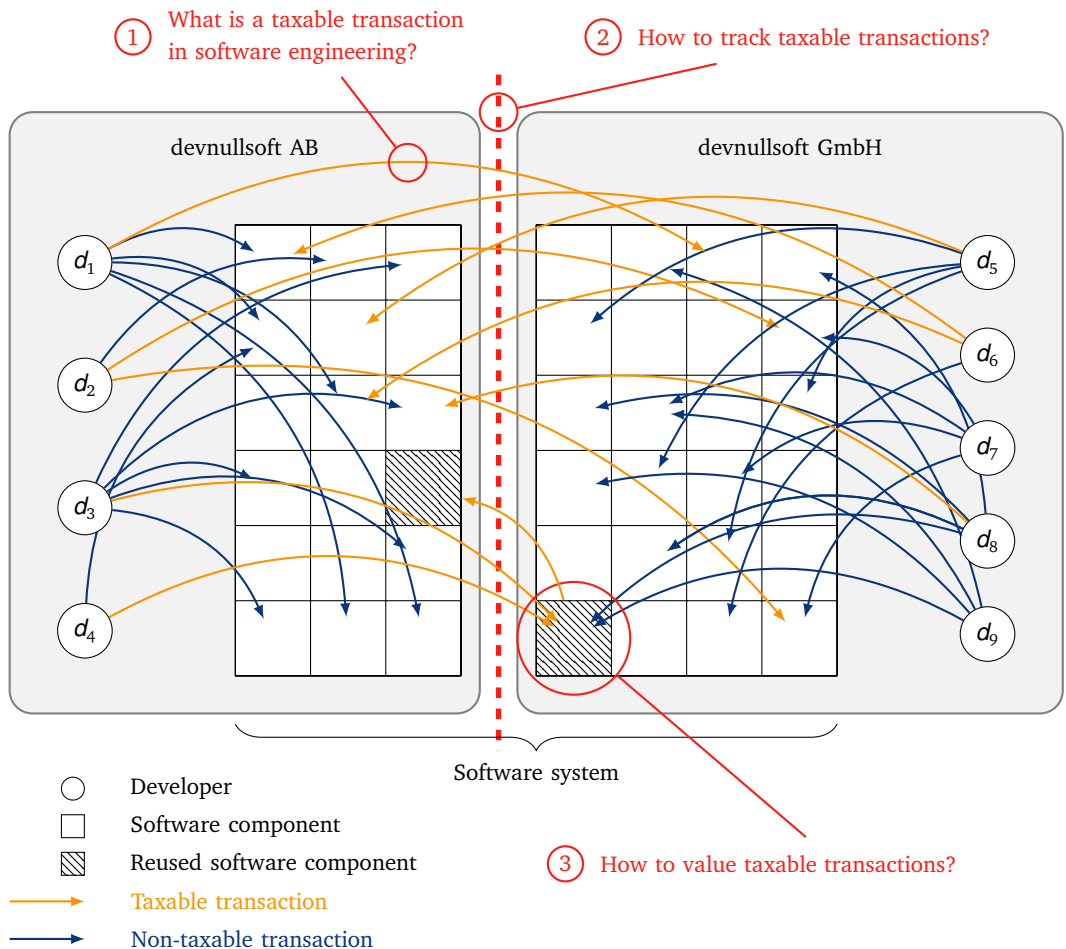

In the following, we discuss three main high-level challenges for transfer pricing in collaborative software engineering within multinational enterprises. Figure~\ref{fig:challenges} highlights the complexity in modern collaborative software engineering at \textit{devnullsoft Group} and where those three challenges apply. 

\subsection*{Challenge 1: What is a taxable transaction in software engineering?}

The trouble for transfer pricing in software engineering begins with a fundamental question: What is actually a taxable transaction in the context of collaborative software engineering? We simply do not know what types or characteristics types or characteristics classify a taxable exchange of intangibles or services across the boundaries of a country in the context of software engineering. 

Among other potentially relevant types of taxable intangibles, such as \emph{goodwill} or \emph{group synergies}, we discuss in this and the following subsections two types of intangibles that are highly relevant for software engineering: know-how and licenses.

The OECD Transfer Pricing Guidelines define \textbf{know-how} as the ``proprietary information or knowledge that assist[s] or improve[s] a commercial activity, but that [is] not registered for protection in the manner of a patent or trademark``. The commercial activity includes the manufacturing, marketing, research and development of and for a software system. 

Does the OECD definition imply that all types of information exchanged during collaborative software engineering are know-how? On the one hand, yes, since all information is proprietary and, to some extent, contributes to the software being developed or its engineering processes. But on the other hand, how do we know which information assists or improves the commercial activity, meaning the engineering of the software system, over time? For example, a quick and dirty bug fix without sufficient documentation or testing may improve the software system in the present but makes changes more costly or even impossible in the future. 
Making such sub-optimal decisions leads to incurring technical debt\autocite{Rios2018}, which is potentially relevant for taxation. Thinking the concept of technical debt ahead, we have begun to understand that similar to physical, tangible assets, software assets degrade and lose value inevitably due to intentional or unintentional decisions caused by technical or non-technical manipulation of the asset or associated assets during all stages of the product life-cycle.\autocite{Zabardast20222} Such an asset degradation will also be of great interest from a taxation perspective. 

The second type of intangibles highly relevant to transfer pricing in software engineering if transferred across borders is \textbf{licenses}. Although maybe not even explicit, the company-internal use and reuse of components is an instance of licensing. Complex software systems are not monolithic blocks of code but consist of components that are developed, shared, and reused by separate teams. However, we lack a common understanding of software components and reuse in software engineering for taxation. Not every component is directly used for or in a software-intensive product, but maybe adds value to the product. For example, a well-engineered CI/CD pipeline accelerates the development cycles and brings new features or bug fixes faster to the customers.\autocite{Fitzgerald2017} Furthermore, it is also not always clear who owns, contributes to, or uses a component within a company, and the roles may even change over time.\autocite{Zabardast20221} In contrast to open source, the reuse is often implicit, lacking a company-wide license agreement that clarifies the responsibility and accountability between component owners and users. Even worse, we do not even know if our definition and understanding of code ownership\autocite{Nordberg2003} suffices the definition of ownership in a taxation context.

Additionally, we see an interplay between those two types of intangibles, know-how and licensing, since they may be two sides of the same collaboration. For example, when code contributions from the component user support instance of reuse. 

\begin{mdframed}[nobreak=true, frametitle={Insight}, skipabove=1em]
Identifying the taxable transactions requires either a holistic perspective of software engineering or at least suitable, practical, and accurate proxies. Compliant software engineering needs a common understanding and a taxonomy of taxable transactions specific to software engineering. 
\end{mdframed}

\subsection*{Challenge 2: How to track cross-border transactions in software engineering?}

Also, the practical tracking of taxable know-how and licensing (and potentially other types of intangibles) is a challenge on its own. 

Tracking know-how is an inherently difficult task. Since the teams collaborating are no longer colocated, numerous tools enable an exchange of know-how in software engineering. Those tools are suitable as rich data sources to different extents: While domain-specific tools like issue trackers or collaborative software development platforms like GitHub or Gitlab often track the exchanges very thoroughly, other communication and collaboration tools do not: Online meetings, for example, can facilitate an exchange of taxable intangibles, but this exchange is not tracked by any tool. But even if there is a rich data basis available, leveraging those data sources is problematic for following reasons: 

\begin{itemize}
\item[{\ieeeguilsinglright}]{\it Establishing location}---%
It can be difficult to establish the location of collaborators or capture when a location of a collaborator has changed, because organizations often preserve only the latest version of the organizational structures.

\item[{\ieeeguilsinglright}]{Privacy}---%
Analyzing the complete communication of developers may be perceived as a measure of surveillance, which raises ethical and legal concerns related to privacy. 
\end{itemize}

In contrast to the potentially rich sources for tracing taxable transactions from collaboration tools used in software engineering, tracking company-internal reuse often lacks a solid data basis. Although companies often track the reuse of external open-source components for open-source license compliance purposes, those tools are rarely used or suitable for tracking company-internal reuse. Also, only reuse that crosses borders is taxable; information that is often not available or stored over time---although component ownership is not static and may be subject to change. 

\begin{mdframed}[nobreak=true, frametitle={Insight}, skipabove=1em, skipbelow=0pt]
Data for tracking taxable transactions may be incomplete, faulty with respect to location, or restricted. There is no dedicated tool support yet for the practical transfer price determination. 
\end{mdframed}

\subsection*{Challenge 3: How to value taxable transactions in software engineering?}

While it is inherently difficult to tax intangibles in general, things are even more complicated in software engineering. Potentially taxable intangibles cover a large range of granularity in software engineering: They may be as large as a microservice providing user authentication used by microservices of other teams ($\rightarrow$ intangible \emph{licenses}) or as small as a code change, code review, or bug report ($\rightarrow$ intangible \emph{know-how}). Although the code change or feedback in a code review is small---maybe even only one line of code, like in the case of the \emph{Heartbleed} security bug in the OpenSSL cryptography library from 2014\autocite{Carvalho2014}---the potential impact on the software system can be tremendous or even fatal. A software change or a code review delivers value through impact, not size. 

The same applies to licensing. The number of use relations of a software component or its size (however defined) does not reflect the value provided to the software-intensive product. While the software component for user authentication may be important for operating the software-intensive product and, therefore, has a large amount of depedent software components, it is not differentiating and may even be considered a commodity. 

This means we cannot simply use purely quantitative measurements for transfer pricing. However, the sheer mass of small, fine-grained transactions of all types makes a human qualitative case-by-case evaluation impossible. 

\begin{mdframed}[nobreak=true, frametitle={Insight}, skipabove=1em, skipbelow=0pt]
A purely quantitative valuing can hardly reflect the value of transactions; however, a purely qualitative assessment does not scale with the magnitude of cross-border transactions in modern software development. 
\end{mdframed}

\section{An industrial example of cross-border collaboration}\label{sec:measurement}

So, is cross-border collaboration and, therefore, also the taxation of it, a real issue? To estimate the prevalence of  cross-border collaboration, we measure cross-border code reviews as proxy for cross-border collaboration in a typical industrial setting. 

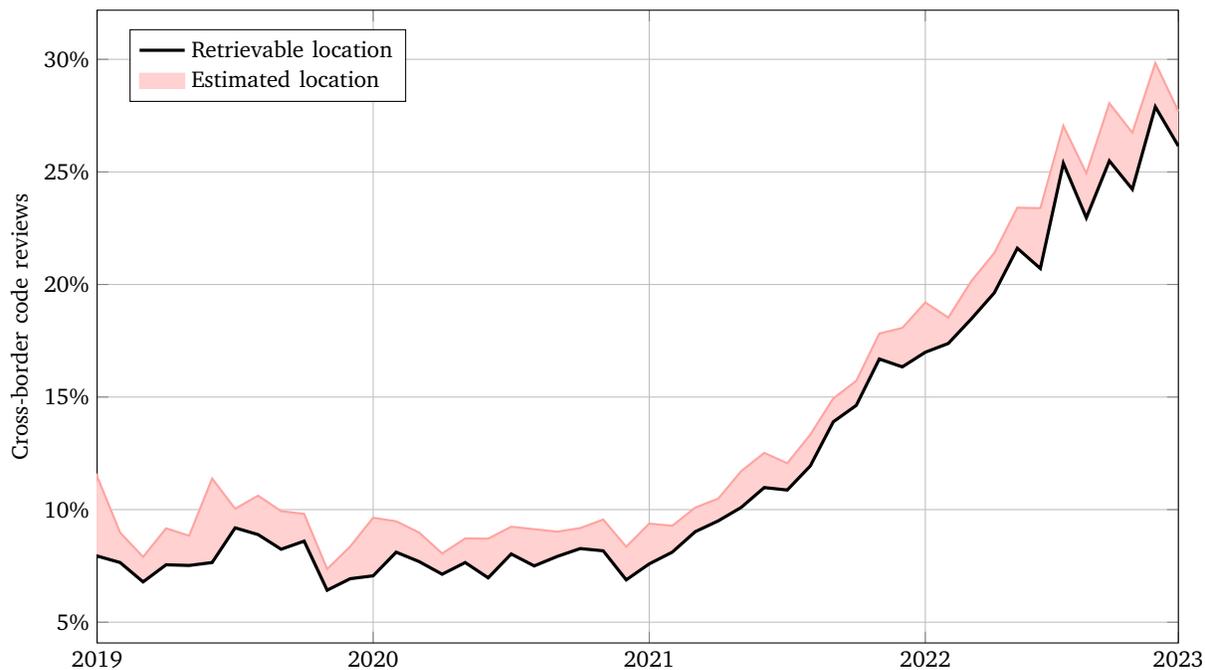
\begin{figure*}[!htp]
\centering
\begin{tikzpicture}
\begin{axis} [
width=\textwidth,
height=10cm, 
legend pos=north west,
legend cell align={left},
grid=both,
enlarge x limits=false,
enlarge y limits=true,
xtick={1, 13, 25, 37, 48},
xticklabels={2019, 2020, 2021, 2022, 2023},
xmin={1},
xmax={48},
clip=true,
ytick={0.05, 0.1, 0.15, 0.2, 0.25, 0.3},
yticklabels={{5\%}, {10\%}, {15\%}, {20\%}, {25\%}, {30\%}},
ylabel={Cross-border code reviews},
]
\addplot[very thick, black, name path=lower] table [x=index, y=lower_crossborder, col sep=comma] {figures/results.csv};
\addplot[thick, ACMRed!40, name path=upper] table [x=index, y=upper_crossborder, col sep=comma] {figures/results.csv};
\addplot[ACMRed!20, area legend,] fill between[of=lower and upper];
\legend{
Retrievable location,
,
Estimated location,
}

\end{axis}
\end{tikzpicture}
\caption{The share of cross-border code review at our case company in the years 2019, 2020, 2021 and 2022 (black line) monthly sampled. Since not all historical locations of all code review participants could be reliably retrieved, the share of cross-border reviews could be more significant (indicated by the red area).} 
\label{fig:measurement}
\end{figure*}

A cross-border code review is code review with participants from more than one country. Although it originated in collocated, waterfall-like code inspections, its modern stances are lightweight and asynchronous discussions among developers around a code change. Different tools are in use, for example, Gerrit or Github and Gitlab with their implementation of code review as so-called pull and merge requests. Although code review is by far not the only type of collaboration that may include taxable transactions and is also likely not sufficient to determine company-wide transfer prices, the following characteristics make code review a suitable first proxy for cross-border collaboration: 

\begin{itemize}
\item[{\ieeeguilsinglright}]{More than code only}---Code review not only includes the actual code change and its authors but also includes the feedback from reviewers that may have formed or changed the code change significantly but is no longer visible in the repository after merging the code change into the code base. Therefore, our proxy goes beyond existing code-based measurements for collaboration.\autocite{Capraro2018}
\item[{\ieeeguilsinglright}]{Accessible \& complete}---The code review discussions are (company-internally) public by default and are, thus, accessible. Unlike other tools like instant messaging services or e-mail, code review does not split into public and private, whose analysis may cause privacy concerns.
\item[{\ieeeguilsinglright}]{Persistent}---Code review tools are the backbone of modern code review and ease data extraction. Other types of code review (for example, private or synchronous discussion around a code change through meetings or instant messaging) may not be captured through the tooling, though. 
\end{itemize}

We measured the share of cross-border code reviews at a multinational company delivering software and related services worldwide with main R\&D locations in three countries. For many years the company has tried to allocate products to particular sites to avoid the burden of cross-border collaboration. However, our analysis shows that developers represent more than 25 locations because the new corporate work flexibility policy permits relocations.\autocite{Smite2022} The company uses a single, central, and company-wide tool for its internal software development and code review. Understandably, our case company wants to remain anonymous. Therefore, we are not able to describe the case any further. However, we believe that our case company is exemplary for a multinational enterprise developing software. 

From the code review tool, we extracted all code reviews that were completed in 2019, 2020, 2021, and 2022 including their activities. All bot activities were removed and were not considered in our analysis. We then modelled code reviews as communication channels among code review participants.\autocite{Dorner2022} We consider a code review as a discussion thread that is completed as soon as no more information regarding a particular code change is exchanged (i.e., the code review is closed). We complement each code review participant with the information of the country of the employing subsidiaries at the time of the code review. 

We provide a replication package to reproduce our results for any GitHub Enterprise instance\footnote{See \url{https://github.com/michaeldorner/tax_se}}. Due to the sensitive topic, we are not able to share our data.

Figure~\ref{fig:measurement} shows an increase in relative cross-border code reviews over time. The share of cross-border code reviews was between 6\% and 10\% in 2019 and 2020. Yet, we see a further steep increase reaching between 25\% and 30\% at the end of 2022. 

Interestingly, 6\% of all cross-border code reviews involve participants from more than two countries. This means transfer pricing in collaborative software engineering becomes not only a bilateral but a multilateral problem with not only two but multiple---in our case company up to six---different jurisdictions and tax authorities involved in the transfer pricing process. 

Although the share of cross-border collaboration may vary between companies, yet, our findings suggest that---through the proxy of cross-border code reviews---cross-border collaboration becomes a significant part of daily life in multinational software companies. It is fair to assume that a further increase in cross-border collaborations in software engineering will draw the attention of tax authorities.

\section{Conclusion} \label{sec:conclusion}

On the one hand, the arm's length principle is the de-facto standard for multinational enterprises that any multinational company must comply with. On the other hand, software engineering is highly collaborative---beyond geographical and organizational boundaries. Determining a reasonable transfer price for this cross-border collaboration brings the general challenge of taxing intangibles to a new level of complexity. 

Pretending to be dead for tax reasons is no option because ignoring the significant cross-border collaboration in modern software development, as we exemplarily found, is a slippery slope: Cross-border collaborations in software engineering will draw the attention of tax authorities. Also, ceasing or forbidding all cross-border collaboration in software engineering is not a valid solution: Reversing the collaborative nature of modern software engineering is likely too costly and takes too long. 

Obviously, neither there are simple solutions for such a complex and interdisciplinary problem, nor a single article can solve this complex problem potentially affecting every software-developing company with developers employed by subsidiaries in more than one country. However, our article aims to bring this eminent and unsolved problem of taxing collaborative software engineering to the audience that can solve this issue. As a software engineering community, we will need to find a common understanding of what constitutes taxable transactions and each company that develops software collaboratively within more than one country needs to learn how to track and value cross-border collaboration, how to estimate the transfer pricing, and how to report these to the tax authorities to be compliant.

\section*{Acknowledgments}

We thank our industry partner for providing the data for this study and interpreting the results. We also thank the anonymous reviewers for their fruitful feedback. This work was supported by the KKS Foundation through the SERT project (Research Profile Grant 2018/010) at Blekinge Institute of Technology.

\printbibliography

\end{document}